# A Conceptual Framework for web Development Projects Based on Project Management and Agile Development Principles

**Martin Tomanek, Radim Cermak and Zdenek Smutny**
**Faculty of Informatics and Statistics, University of Economics, Prague, Czech Republic**
martin.tomanek@vse.cz
radim.cermak@vse.cz
zdenek.smutny@vse.cz

**Abstract:** Companies implement different frameworks and best practices with the objective to improve the project management success rate and improve the business adaptability to the changing business environment. Project management framework (PRINCE2) and agile development framework (Scrum) proved in many cases that they can meet these objectives. However, both frameworks are based on different principles and the use of both frameworks together should be carefully considered. A large amount of money and effort has been invested by companies into establishing their project management environment and processes that follow the classical phased approach where requirements are defined upfront and fixed. But companies want to react more quickly to new global challenges and to the changing business environment. These business requirements then result in the failure of many running projects. Therefore there is a need to enhance the current project management environment so that it is more agile and adoptive to changes. The objective of this paper is to create a conceptual framework that aggregates principles and processes from both frameworks (PRINCE2 and Scrum) with emphasis on their use in web development projects. This paper will discuss the advantages and disadvantages of using the two abovementioned frameworks. Different groups of readers can benefit from the results of this paper. It will help corporate management to decide how a company should set up its own specific framework for managing agile product development projects. Project managers will have a better understanding of agile development principles and how it fits in the classic project management framework. Last but not least, it will help product developers to work in more agile ways and survive in the controlled and complex project environment.

## 1. Introduction and methodology

In 1970 the waterfall development model was introduced by Winston W. Royce. This model has since been used to manage IT projects. This model follows a phased approach, in which the requirements are defined upfront, then the solution is designed, coded, tested and released to production. In many cases this approach failed but of course in many cases it was successful. The project management methods were aligned with this phased approach. For project managers it was easier to manage individual phases according to a plan. Project management frameworks like PMBOK or PRINCE2 were introduced and many project managers trained and certified.

Current business environment demands shorter time to market and greater flexibility of the ever changing business requirements. (Kalina et al., 2013; Basl and Doucek, 2012) Some leaders in the software development field realized that robust and heavy-weight process frameworks simply do not work as expected. They introduced the agile manifesto (Beck et al., 2001) which is followed by many software developers. As a result introducing the agile manifesto, several agile development frameworks were introduced with different scope and focus (Scrum, Extreme Programming, FDD and others).

The chaos report from last year (The Standish Group, 2013) measuring project success rate indicates that 39% of projects were delivered successfully, 18% completely failed and 43% were challenged. The project success rate was slightly improved compared to year 2004 but the percentage of failed projects still remains almost the same.

A shift to agile methods can increase the success rate and mitigate some issues that are typical for heavy-weight methods. Good examples can be found in the case studies (Balada, 2013; Raithatha, 2007), where agile methods were successfully used in large complex software development projects. On the other hand, in the case of long-term projects, there are problems at the level of agile teams managed by a project manager that have an impact on the decision-making process (McAvoy, Butler, 2009). At the level of project management in large companies (with complex IT landscape), we can find other issues in the co-existence of agile methods and plan-driven development in many organizations (Waardenburg and Vliet, 2013). In every way, the current challenge is cooperation between these different approaches, which are applied mainly in large organizations.

This paper will analyse the two most popular and widely used frameworks for project management and agile development. The first one is PRINCE2 and it stands for "Projects IN Controlled Environments" version 2 (Office of Government Commerce, 2009). It has been developed by the UK government agency Office of Government Commerce (OGC) and is mostly used by European countries. The second framework in the scope of this paper is Scrum and it represents the agile product development framework. Scrum was developed by two authors and is described in the Scrum Guide (Schwaber and Sutherland, 2013). This framework is one of the most frequently used agile development frameworks in the world.

The objective of this paper is to propose a new framework that will aggregate processes and principles from the two abovementioned frameworks into a new conceptual framework. Advantages and disadvantages of this framework will be discussed as well. This contribution is conceived as design research, the result of which is an artefact. In our case, it is a conceptual model that will have to be validated by comparative case study (it is inductive approach to the validation of output), see – (Yin, 2008, pp 3-5). This case study is currently planned, but we would like stirred up discussion on this topic. In addition to the basic model we want to mainly discuss different views on the usefulness of such an approach. In this case, we focus on the development of web applications and for this purpose we present a theoretical comparison in the final part of the paper.

## 2. PRINCE2

PRINCE2 is a project management framework that can be used for managing any kind of projects regardless of project scale, type, organization, geography or culture. This broad applicability is caused by the seven underlying guiding principles that create the core of this framework. The principles are: the project has a continued business justification, project roles and responsibilities are defined, the project is managed by stages, the project is managed by exception, the focus is on the delivery of a quality product, the project team learns from experience and the project management framework is tailored to suit its specific project environment and characteristics. Additionally the PRINCE2 framework consists of seven themes and seven processes. Themes represent areas that the project manager should continuously consider during the project lifecycle. The themes are: business case, organization, quality, plans, risk, change and progress.

PRINCE2 is a process-oriented framework. It contains seven processes that guide project stakeholders to direct, manage and deliver the project successfully. The processes provide a guideline to who should do what, when and why. The process diagram is depicted in Figure 1 PRINCE2 Process Diagram. The first process called "*Starting up a project*" has the objective to make sure that the business case is drafted and the project manager is appointed, see (Juricek, 2014). After this process is completed then the project board should decide to initiate the project. Project board's decisions are part of the process called "*Directing a project*". When the project is authorised to continue then the project manager prepares different strategies for managing risks, quality, configuration and communication. The project manager also updates the business case and prepares the project plan. These activities are executed in the process called *"Initiating a project"*. The project manager in the next process called "*Controlling a stage*" assigns work to delivery units, monitors progress and solves issues related to the assigned work. Finally there is the process called "*Managing product delivery*" in which delivery teams accept, execute and deliver work. The following process, "*Managing a stage boundary*", is another process designed mainly for the project board, in which the project board receives information about current progress of the project. The last process, "*Closing a project*", has the objective to verify that the final product meets the requirements and to hand the product over to the product owner.

The last area important for this paper is roles and responsibilities. The project board has overall responsibility for the project success. The project board consists of executive, senior user and senior supplier. The project manager is primarily responsible for ensuring that the project creates the required products or outcomes. The team manager is responsible for the product delivery in required quality, timescale and costs.

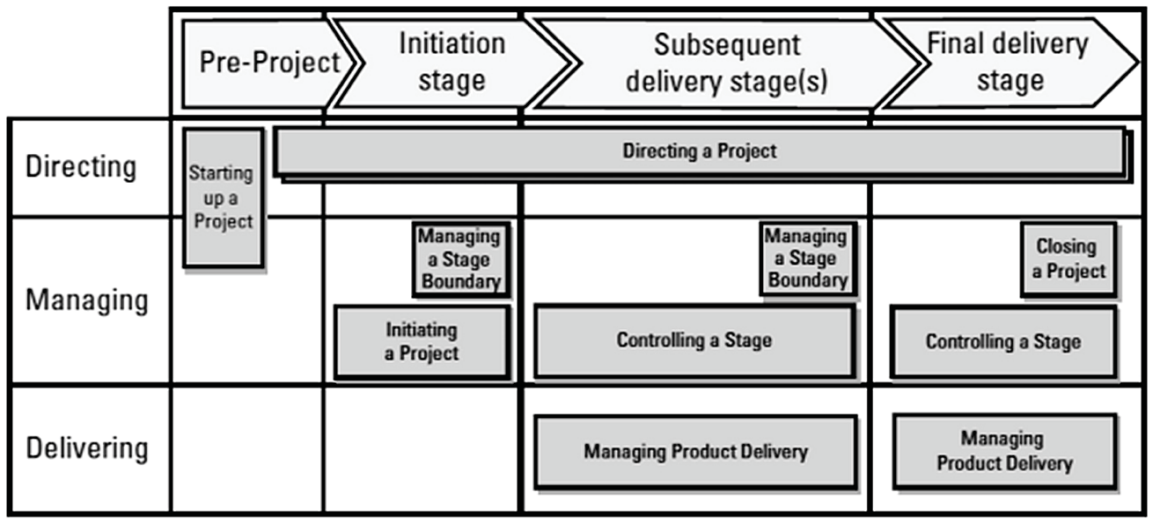

**Figure 1:** PRINCE2 process diagram. Source: (Office of Government Commerce, 2009)

## 3. Scrum

Scrum is an agile product development framework that is mainly used for iterative and incremental software development. Scrum as a framework is light-weight compared to PRINCE2. The core of Scrum framework is that customer requirements can be changed during development and the product should be developed iteratively. Iterations are called sprints and every sprint starts with a sprint planning meeting where the customer reviews and prioritizes requirements. Then the development team works together to develop product features and deliver a shippable product by the end of each sprint. This shippable increment or product is presented in the sprint review meeting where the customer can see the product and think about further development. The most frequent meetings are daily stand-up meetings where development team members discuss what they have done since the last meeting, what they will do in the coming days and whether they face any impediments.

Scrum defines three roles. The Product Owner represents customers and is responsible for defining and prioritizing product requirements and records them in the product backlog. The Development Team is responsible for delivering the potentially shippable product by the end of each sprint. The Scrum Master facilitates Scrum meetings and ensures the development team can work as efficiently as possible.

## 4. Conceptual framework

The conceptual framework introduced by this paper consists of the alignment of principles and processes.

In the previous chapter the principles of both frameworks were discussed. Now this paper will focus on the alignment of project management principles to the agile development approach. All the seven PRINCE2 principles will be discussed in terms of how they are affected by the agile development approach defined by Scrum.

As both frameworks are process oriented it allows us to create a process model that integrates both frameworks. This process model shows the major integration points and the process flow. The integrated process model can be found below.

### 4.1 Alignment of project management principles with agile development approach

*Continued business justification*

The project is initiated and run with the objective to deliver the project outcomes. These outcomes should bring benefits to business. If there is no value in the outcomes for business then the project has no justification for being continued. Therefore there should be a clear statement defining the business justification. The justification is described and approved in the business case. This justification should be verified by the product

*Martin Tomanek, Radim Cermak and Zdenek Smutny*

owner and the business case updated, ideally after each sprint. This frequent review is supported by every increment that helps to identify projects with no business justification.

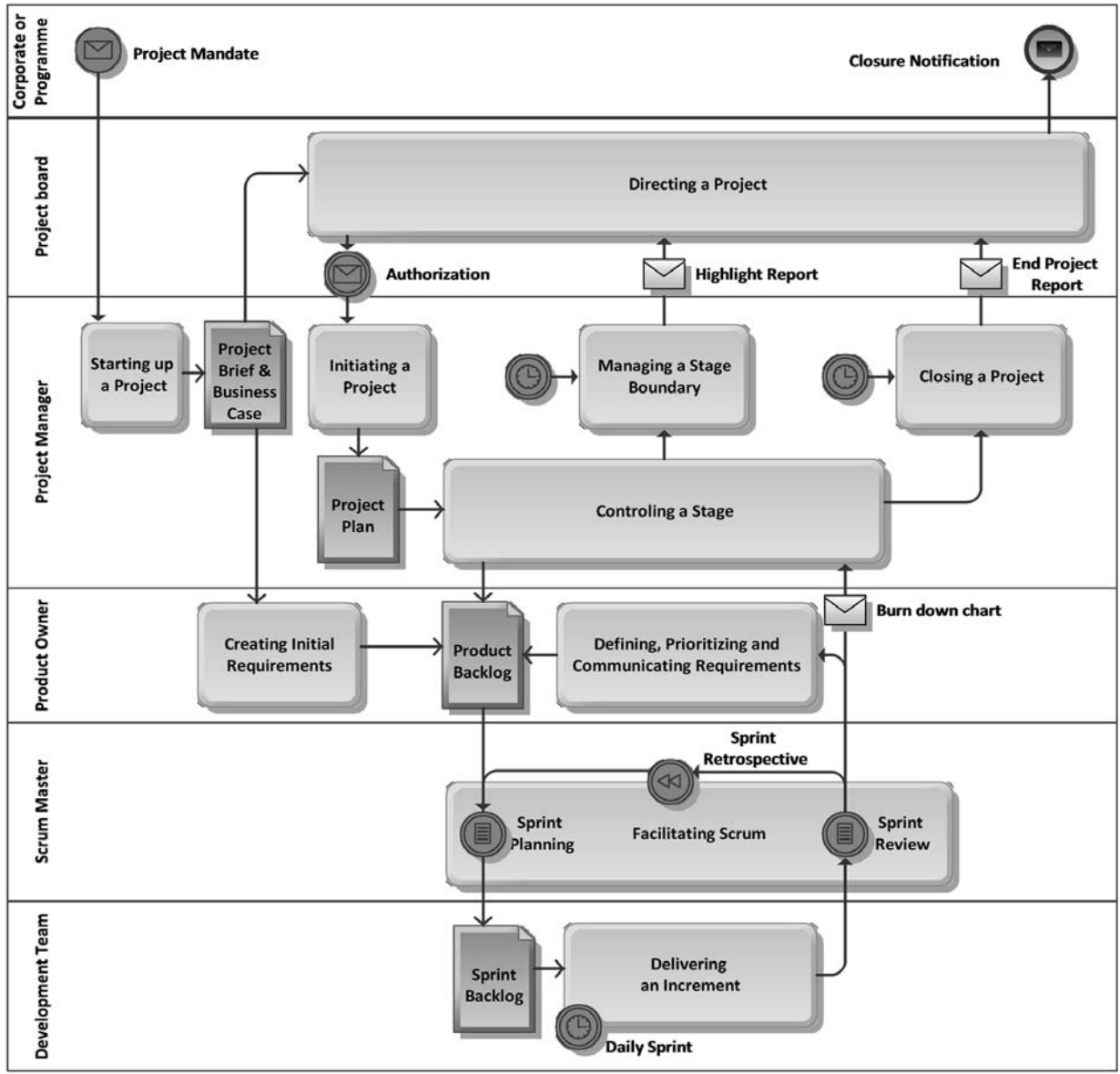

**Figure 2:** Process model integrating PRINCE2 and Scrum framework. Source: Authors.

*Learn from experience*

Both frameworks contain mechanisms for learning from experience. Scrum recommends running sprint retrospective events where the whole Scrum team can review the existing process and plan how to improve it. PRINCE2 recommends using a lessons log where the project lessons can be recorded and used later in the project or by other projects. This lessons log can be used also by the development team. In the closing phase of the project the project manager should ensure that the lessons report is generated and can be shared by other projects or teams using Scrum.

*Defined roles and responsibilities*

Both frameworks define roles and their responsibilities. The most important project role defined by PRINCE2 is the project manager who is responsible for project delivery. The project board consisting of executive, senior supplier and senior user is accountable for project outcomes and benefits. The last major role is the team leader who is responsible for delivering the accepted work packages. Scrum defines the scrum team as a self-organizing team consisting of development team, product owner and scrum master. Two roles from each framework are similar and can be executed by one person in small projects. The first similar pair of roles is the product owner and the senior user. Both roles represent customer view on the project and the product. The second similar pair of roles is the team leader and the scrum master. If the team leader is an experienced scrum master then this can be a good combination of two roles into one person.

*Manage by stages*

The project is split into stages that help the project manager to review the current progress and initiate necessary actions when required. The managing by stages principle also provides opportunities for the project board to review the project status and provide decisions that are outside the project manager responsibility. PRINCE2 stages can be easily aligned and synchronized with sprints defined by Scrum. Sprint should take a month at maximum and the stages can be planned accordingly. Synchronization of stages and sprints helps to create valuable highlight reports with up-to-date information for the project board and other project stakeholders.

*Manage by exception*

The project manager manages the project and tries to keep the project within its budget, deliver the project on time and in required quality. For these aspects the project manager has defined boundaries within which he or she can act. If these limits are broken then project board involvement is required. Scrum is split between time-limited sprints that mitigate the risk of the project being late or overspending. The product and its increments are reviewed on a regular basis and this helps to improve the quality and to identify and fix errors earlier in the development phase. Frequent delivery of increments helps to plan and forecast the needed budget because the development cost of each sprint is similar and predictable.

*Focus on products*

Both PRINCE2 and Scrum frameworks aim to deliver quality products. PRINCE2 tends to describe the product in detail including requirements and acceptance criteria prior to product development. Scrum is focused more on the evolution of requirements and continuous definition and prioritization. If we combine these two approaches then we can recommend defining the high-level critical requirements during the initiating phase and record them in the business case. These high-level requirements should be considered during the discussion about business justification and used as the basis for a feasibility study. After the business case is approved and the project initiated then these requirements should be analysed and worked out in greater detail. Additional emerging requirements can be added to the product backlog at any time. Then during the sprint planning the most value-adding or the most risky requirements are selected to be developed first and are included in the sprint.

*Tailor to suit the project environment*

Tailoring of heavy-weight PRINCE2 framework is required by PRINCE2 by default. Scrum as the product development framework is a good complement to the pure project management framework. Scrum can fill in the missing details in the PRINCE2 process "managing a product delivery". After combining these two frameworks the new framework can be created that focuses on project management of agile product development.

### 4.2 Integrated process model

The Scrum development process fits into the PRINCE2 process framework. Scrum as the development framework can replace the managing delivery process defined by PRINCE2. In this way PRINCE2 wraps the delivery process of the project products. The benefit of Scrum is that it contains guidance on how to develop the product in an efficient way and adapt to the changing environment.

The project manager starts the project as any other project by putting together the requirements and expected costs. If the product owner exists in the current line organization then the project manager should cooperate with him or her to validate and clarify the requirements. The project manager also designs and appoints the project management team. Expected project benefits are summarized in the business case and together with estimated costs presented to the project board. The project board can then decide if the project should be initiated and authorize project execution.

When the project can be started, the project manager plans the following steps in greater detail. The project manager plans the stages and related sprints. The product owner in the meantime should move the initial requirements from the business case and transfer them into the product backlog. The scrum master should be appointed and together with the project manager should facilitate the first sprint and to have the first sprint planning event.

The product owner with the development team discusses the product backlog and decides what requirements can be delivered in the next sprint. All selected requirements are transferred to the sprint backlog and described in more detail by the development team. When the sprint backlog is created, the team starts designing, developing and testing the product increment. They meet every day and during 15 minutes they discuss what has been done, what will be accomplished soon and whether they face any impediments. These daily stand-up meetings should be facilitated by the scrum master and if the product owner is available then the product owner should attend these daily sprints as well.

When the delivery part of scrum is over, the development team demonstrates the product increment to the product owner in the sprint review meeting. The product owner can decide if the product increment should be released to production or further developed. After the sprint review meeting, the product owner can explore the product increment, create new requirements, update the existing ones and prioritize all of them in the product backlog.

The project manager after the sprint review meeting should update the highlight report where he or she summarizes the progress of the project. This report is then presented to the project board to inform them of the project status. Scrum also defines a more frequent report called the burn-down chart that contains the summary of completed tasks and estimated time for completing all the remaining requirements. This chart helps the project manager to have a better operational overview of requirements to be delivered in the sprint.

## 5. The use of conceptual framework in the field of web development

Web development is constantly on the rise along with the use of internet-based technologies which are positively accepted by society. The most of contemporary ecosystems of operating systems (on different devices) implicitly supports the web browsing option and therefore the use of web applications. Several years ago, the area of web development encompassed mainly small-scale projects, but currently there is a growing demand for the development of specific web applications of the extent of medium and large projects (or a larger number of smaller projects) built on various web technologies. This is caused by the efforts of primarily commercial subjects to reflect their activities in the environment of services on the internet (e.g. banks, insurance companies, retail companies), but also by a gradual transition from platform-dependent solutions to the more universal web-based applications (e.g. GPS navigation). This transition entails (on the side of web solution itself) some advantages as well as disadvantages:

- *Advantages of web applications:* available 24 hours a day, 7 days a week; zero install (only web browser needed); they can reach anybody anywhere in the world; centralised data is secure and easy to backup; quick and easy updates (always up-to-date); low spec. mobiles, computers or tablets can be used;

- *Disadvantages of web applications:* slower (run over web browser); internet is not always available; greater complexity means longer development (mix of many technologies as HTML5, CSS3, PHP, JS); various standards supported in various browsers; security risks.

The environment of services on the internet fluctuates a lot; this is why the development of these applications has to be adapted to the agile approach. On the other hand, medium or large projects (in scope) created by variously large teams (according to the – current – financial possibilities of the client) carry with them the need for effective management of the whole project (or partial smaller projects). The conceptual framework presented here links the elements of PRINCE2 and Scrum methodologies and provides the required superstructure of the development approach in the form of project management and control at a higher level (project manager and board). It also allows the integration of this conceptual framework into larger organizational structures in large companies, offering better management of individual modules (parts or branches with different variants) of the project, including the possibility of effective management of otherwise separate developments. This concerns primarily the improvement of planning and control mechanisms at a higher level of project management. This is important from the point of view that these projects can be long-term, so the software is subject to continuous development (paid for its use and not for a final version handed over to the client). For these reasons, companies engaged in business via internet make long-term portfolios of different sized projects that are actively developed according to the needs and possibilities of the clients.

The following table compares the two main approaches to software development and compares their main features with the self-designed conceptual framework, in which an extension of agile development was added in the form of more robust project management. This allows particularly a more efficient management of medium and large projects, or a larger portfolio of smaller projects. The advantage of two levels of management (agile at the development level compared to the more rigid control at a higher level of the project) lies in the ability for rapid change, but also in following the predetermined direction, so that the project is able to meet the business objectives defined at the beginning of the project (including its successful completion). In Table 1, we highlight in bold type the important changes as compared with net agile development.

**Table 1:** Differences between traditional development, agile development and agile development with project management: Source: Authors, based on (Stoica et al., 2013)

|  | **Traditional development (e.g. Rational Unified Process)** | **Agile development (e.g. Scrum)** | **Agile development with project management (Scrum and PRINCE2)** |
|---|---|---|---|
| **Management style** | Command and control | Leadership and collaboration | Leadership, collaboration, and control |
| **Knowledge management** | Explicit | Tacit | Tacit (explicit if required) |
| **Communication** | Formal | Informal | Informal (development team), formal (project board), mixed (project team) |
| **Development model** | Life cycle model (waterfall, spiral or modified models) | Evolutionary-delivery model | Evolutionary-delivery model |
| **Organizational structure** | Mechanic (bureaucratic, high formalization), targeting large organization | Organic (flexible and participative, encourages social cooperation), targeting small and medium organizations | Organic, targeting especially small and medium organizations. But thanks to the robustness of PRINCE2, it can be used also by large organizations. |
| **Quality control** | Difficult planning and strict control. Difficult and late testing | Permanent control of requirements, design and solutions. Permanent testing. | Permanent control of requirements, design and solutions. Permanent testing and bug fixing. |
| **User requirements** | Detailed and defined before coding/implementation | Interactive input | Interactive input (high-level requirements can be defined upfront) |
| **Cost of restart** | High | Low | Medium - Low |
| **Development direction** | Fixed | Easily changeable | Easily changeable in defined limits |
| **Testing** | After coding is completed | Every iteration | Every iteration |
| **Client involvement** | Low | High | High |
| **Additional abilities required from developers** | Nothing in particular | Interpersonal abilities and basic knowledge of the business | Interpersonal abilities and basic knowledge of the business |
| **Appropriate scale of the project** | Large scale | Low and medium scale | Medium and large scale (low also possible) |
| **Requirements** | Very stable, known in advance | Emergent, with rapid changes | Emergent, with rapid changes. High-level requirements known in advance |
| **Architecture** | Design for current and predictable requirements | Design for current requirements | Design for current requirements, blueprint for high-level requirements |
| **Remodelling** | Expensive | Not expensive | Not expensive at the level of |

|  | **Traditional development (e.g. Rational Unified Process)** | **Agile development (e.g. Scrum)** | **Agile development with project management (Scrum and PRINCE2)** |
|---|---|---|---|
|  |  |  | one module |
| **Size** | Large teams and projects | Small teams and projects | Small teams and small, medium or large projects |
| **Primary objectives** | High safety | Quick value | Quick value and high safety of project as a whole |

The table clearly shows that the combination of agile development and more rigid management is easily applicable in particular to the development of medium and large web projects, where the use of the proposed framework brings many advantages. In the field of web application development small scale projects are still prevalent; those include common corporate presentations, small e-shops or professional portals. These projects too need high quality management with defined objectives and control during the development of the application. Therefore, the proposed conceptual framework is suitable for this type of projects (including a broader portfolio of small projects). However, it is important to keep in mind certain distinctive characteristics of smaller projects, which must be reflected in the way the chosen management style is used/applied.

The development of small web applications compared to larger-scale projects is different in the extent of the complexity of the proposed applications, the size of the team and therefore the number of roles that can be identified in the project. It is typical for smaller projects to merge certain roles in one person. In a typical small-scale project, there are in fact only two main roles: Team Leader (combining the role of Scrum Master and constitutes the essence of the team, which is very small or consists of external members, who are hired, if necessary) and Product Owner, who represents the customer. The lower level of project complexity is reflected in the number of sprints. The minimum number is three sprints: the first is focused on component layout and design, the second on creating key functionality of web applications and the third on the control of the beta version of the website (a version that is finished in terms of functionality and design, but there is a possibility of minor deficiencies). This minimalist approach to the number of sprints has another consequence, which is a reduced capacity for changes in functionality (due to the smaller amount of sprints) and that is connected with higher requirements on the description of user requirements. For these reasons, we recommend a higher number of smaller sprints (e.g. weekly), if it is allowed by the timetable for the supply of the final version of the web application (projects longer than one month). The need to describe user requirements is also important in terms of the budget for the project, which is mostly fixed, based on the initial project specification. In the light of those specificities it seems appropriate to apply our proposed framework also to developing web applications in smaller scale.

## 5.1 Advantages and disadvantages of the new conceptual framework

The following potential advantages and disadvantages of the new framework should be considered when applying the results of this paper in order to improve the web development project success rate.

*Advantages of this framework for web development projects:*

- It supports the integration of Scrum into the more complex organizational structures.
- It supports the management of medium to large projects or a portfolio of small projects.
- It supports the management of different development teams using Scrum (Scrum-of-Scrum)
- It provides two management layers. The first one is focused on development efficiency and the second is focused on keeping the project within budget, quality standard and on time.

*Disadvantages of this framework for web development projects:*

- Web development projects can be so small and short-lived that no formal framework is needed and no portfolio management is applied.
- The customer and the development team must be committed to the project otherwise the benefits of the framework will not be fully realized.
- When the user requirements are stable and can be defined upfront then this framework is less effective than the traditional approach.

## 6. Conclusion

The introductory section of the paper presented the importance of the currently applied approaches to software development with the assistance of agile approach as well as good project management with emphasis on ensuring a successful completion of the project. We also briefly outlined the basic principles of the two methodologies – the first focused on agile development (Scrum) and the second on project management (PRINCE2) – which we had decided to use to design our own conceptual framework linking (unifying) these two methodologies. The resulting conceptual framework provides a new qualitative value, connecting the strengths of both methodologies in order to ensure effective development and also strict compliance with project goals. Finally, we discuss the possibilities and applicability of the conceptual framework for developing web applications including comparison with the traditional and the purely agile way of development. One of the main benefits of this conceptual framework is that it can help the customers to see the more frequent product increments and determine whether the project still has business justification. More frequent customer involvement can help the development team to understand the requirements in greater detail.

In order to fully examine the conceptual framework and to support higher value creation the following process artefacts need to be described in more detail. The individual processes should be described especially in terms of process objectives, description, inputs, outputs and RACI (responsible, accountable, consulted and informed roles). The roles from both frameworks should be also aligned and their responsibilities clearly defined to support the process execution. Last but not least, the process templates should be created to guide the process practitioners to easily follow the integrated process model. The research limitation of this paper is exclusion of people skills and motivation. Only principles and process models were considered but the people factor is the most crucial in any software development projects. Scrum supports the collaborative culture and PRINCE2 is mostly used as the command and control approach. These two different approaches may decrease the real benefits of this new framework and therefore the combination of the mentioned two frameworks should be carefully understood and considered.

## Acknowledgements

This paper was prepared thanks to the IGA grant VSE IGS F4/5/2013.

## References

Balada, J. (2013) "Scrum Adoption for Information System Development within Complex Environments", *39th International Conference on Current Trends in Theory and Practice of Computer Science*, Spindleruv Mlyn, pp 42-53.

Basl, J. and Doucek, P. (2012) "ICT and Innovations in context of the sustainable development in Europe", *20th Interdisciplinary Information Management Talks*, Jindrichuv Hradec, pp 153-161.

Beck, K., Beedle, M., Bennekum, A. van, Cockburn, A., Cunningham, W., Fowler, M., Grenning, J., Highsmith, J., Hunt, A., Jeffries, R., Kern, J., Marick, B., C. Martin, R., Mellor, S., Schwaber, K., Sutherland, J. and Thomas, D. (2001) "Manifesto for Agile Software Development", [online], Ward Cunningham, http://www.agilemanifesto.org.

Juricek, J. (2014) "Business Case in terms of business management and IT Governance", *Acta Informatica Pragensia*, Vol 3, No. 1, pp 33-43.

Kalina, J., Smutny, Z. and Reznicek, V. (2013) "Business Process Maturity as a Case of Managerial Cybernetics and Effective Information Management", *7th European Conference on Information Management and Evaluation*, Gdansk, pp 215-221.

McAvoy, J. and Butler, T. (2009) "The role of project management in ineffective decision making within Agile software development projects", *European Journal of Information Systems*, Vol 18, No. 4, pp 372-383.

Office of Government Commerce (2009) *An introduction to PRINCE2: managing and directing successful projects*, Stationery Office Books, London.

Raithatha, D. (2007) "Making the whole product agile - A product owners perspective", *8th International Conference on Agile Processes in Software Engineering and Extreme Programming*, Como, pp 184-187.

Schwaber, K. and Sutherland, J. (2013) "The definitive guide to Scrum: The rules of the game", [online], Scrum.org, https://www.scrum.org/Portals/0/Documents/Scrum%20Guides/Scrum_Guide.pdf.

Stoica, M., Mircea, M. and Ghilic-Micu, B. (2013) "Software Development: Agile vs. Traditional", *Informatica Economica*, Vol 18, No. 4, pp 64-76.

The Standish Group (2013) "CHAOS Manifesto 2013 – Think Big, Act Small", [online], The Standish Group International, http://www.versionone.com/assets/img/files/CHAOSManifesto2013.pdf.

van Waardenburg, G. and van Vliet, H. (2013) "When agile meets the enterprise", *Information and software technology*, Vol 55, No. 12, pp 2154-2171.

Yin, R. K. (2008) *Case Study Research: Design and Methods*, SAGE Publications, London.